# GRB Probes of the High-z Universe with *EXIST*

Jonathan Grindlay[a] and the *EXIST* Team[b]

[a]*Harvard-Smithsonian Center for Astrophysics*
*60 Garden St.*
*Cambridge, MA 02138*

[b]*Working Group Chairs listed at:* http://EXIST.gsfc.nasa.gov/wgs.html

**Abstract.** The Energetic X-ray Imaging Survey Telescope (EXIST) mission concept was selected for further study under the Astrophysics Strategic Mission Concept Study (ASMCS) program. The mission design is optimized for study of high-z GRBs as probes of the early Universe. With a High Energy Telescope (HET) incorporating a 4.5m$^2$ 5-600keV (CZT; 0.6mm pixels) detector plane for coded aperture imaging a 90$^o$ x 70$^o$ (>10% coding fraction) field of view with 2' resolution and <20" (90% conf.) positions for >5 sigma sources, EXIST will perform rapid (<200sec) slews onto GRBs. Prompt images and spectra are obtained with a co-aligned soft X-ray telescope (SXI; 0.1 - 10keV) and with a 1.1m optical-IR telescope (IRT) simultaneously in 4 bands (0.3 - 0.52μm, 0.52 - 0.9 μm, 0.9 - 1.38 μm, and 1.38 - 2.3 μm). An initial image (100s) will yield prompt identification within the HET error circle from a <2" prompt SXI position; or from VIS vs. IR dropouts or variability. An autonomous spacecraft re-point (<30") will then (at ~300sec after GRB trigger) put the GRB on a 0.3" x 4" slit for either R = 3000 (for AB <21) or R =30 (for AB ~21-25) prompt spectra over the 0.3 - 0.9 μm and 0.9 - 2.3 μm bands. Ambiguous initial identifications will have objective prism spectra for all objects in a 3.75' x 0.75' portion of the 5' x 5' IRT field. This will provide onboard redshifts within ~500-2000sec for most GRBs, reaching z ~20 (for Lyman-α breaks) if such GRBs exist, and spectra for studies of the host galaxy and local re-ionization patchiness as well as intervening cosmic structure. With ~600 GRBs/yr expected, including ~7-10% expected at z >7, EXIST will open a new era in studies of the early Universe as well as carry out a rich program of AGN and transient-source science. An overview of the GRB science objectives and a brief discussion of the overall mission design and operations is given, and example high-z GRB IRT spectra are shown. EXIST is being proposed to the Astro2010 Decadal Survey as a 5 year Medium Class mission that could be launched as early as 2017.

**Keywords:** Gamma-ray Bursts, Early Universe, Large Scale Structure, Black Holes, Jets, Space telescopes
**PACS:** 95.55.-n, -.Ka, -.Fw; 95.75.Fg; 95.85.Kr, -.Ls, .Nv, -.Pw; 97.20.Wt, -.Lf; 98.62.-g; 98.70.Rz

## INTRODUCTION

The most luminous objects known in the observable Universe are Gamma-ray Bursts (GRBs). Many papers (e.g., Bromm and Loeb 2007, hereafter BL07, and references therein) have pointed out the key questions for understanding the early (and later) Universe that can be addressed by using GRBs. These include:

i) the use of GRBs to trace the history of the epoch of reionization (EOR) by studies of the damped Lyα absorption (DLA) profile redward of the line (McQuinn et al 2009);

ii) the possibility of the *unique* detection of individual Pop III stars, the first to form in the Universe, by the GRBs these massive (~100 M$_\odot$) stars may produce upon collapse to stellar mass black holes (SBHs) (BL07);

iii) the star formation history of the early Universe after Pop III and the second (and third) generation stars (Hartmann et al 2009); and the increase of metallicity in the Universe with cosmic time, Z(z), as measured by absorption line presence and (approximate) strengths in the host and intervening galaxies as a function of redshift; and

iv) using GRBs as triggers for gravitational wave detection of (relatively nearby) merging neutron star of black hole binaries, the likely progenitors of short GRBs (Bloom et al 2009).

Other fundamental measures are possible if a significant sample of high-z GRBs can be obtained, but we shall focus on those just listed as primary motivations for *EXIST* studies of GRBs. We shall also briefly mention other new studies of GRBs that the proposed *EXIST* concept would enable – several for the first time. These include high time resolution studies of GRB spectra, which may constrain emission models, and measurements of polarization vs. time and energy in the prompt emission of GRBs, which would give new constraints for both emission and jet models.

This paper is organized as follows. First, a broad overview of the *EXIST* mission is provided since this has changed from the mission concept previously discussed for GRBs (Grindlay 2006) that was studied for the 2006-2008 review by the Beyond Einstein Program Assessment Committee (BEPAC). We then describe the three Instruments proposed for *EXIST*: the large-area High Energy Telescope (HET), a 1.1m optical–nearIR telescope (IRT), and the Soft X-ray Imager (SXI; contributed by Italy/ASI), and how they are combined for GRB studies. The overall mission design and plan is then described, followed by a brief description of the other key science objectives for *EXIST* and how they are coordinated with GRB studies in the scanning vs. pointed phase of the mission. Finally, some example IRT simulated spectra of GRBs are given to illustrate what is possible to *EXIST*.

## OVERVIEW OF *EXIST* MISSION

*EXIST* is first and foremost a hard X-ray imaging (5-600 keV) survey mission to study black holes on all scales. It surveys nearly the full sky every 3h, or every 2 orbits in its low earth orbit (600km and ~15° inclination). The wide-field imager, HET, is a coded aperture telescope with a 90° x 70° (at 10% coding fraction) field of view and 2' resolution, yielding ≤20" source positions for ≥5σ source detections. The first 2y of the *EXIST* mission are devoted to a repeated all-sky survey for which a scanning sky coverage is optimal for both spatial and temporal coverage of black holes. Since the two principal scientific objectives are studies of high-z GRBs and obscured and dormant AGN, the mission now includes a pointed instrument complement to provide enable precise (<1-2") positions and thus unambiguous identifications (in most cases) as well as broad-band (NIR-optical-soft/medium X-ray) spectral coverage. The IRT is a 1.1m telescope, adapted from the commercial *NextView/GeoEye* telescope/mission launched in September 2008 and now providing high quality commercial imaging (optical). The IRT would enable *simultaneous* optical – NIR imaging or spectroscopy covering the entire region 0.3μm – 2.2μm in four bands (see Table 1). The sensitivity is enhanced over ground-based telescopes in all bands due to the usual advantages of a space telescope, but particularly in the NIR where by passively cooling the primary/secondary mirror to (only) -30C, backgrounds a factor of ~$10^3$ lower at 2μm than on the ground are achieved, allowing the IRT to be significantly more sensitive than Keck in the NIR and able to image (10σ) magnitudes AB ~24 in just ~100sec. At the same time, in pointed mode, the SXI (proposed to be contributed by ASI/Italy; Tagliaferri et al. 2009) provides 0.1 – 10 keV imaging and spectroscopy with nearly 8X the sensitivity of the XRT on Swift (from the same Italian Team). With a ~20' FoV and ~15" resolution, the SXI will enable prompt measures of GRB afterglow positions (<2"), variability and spectra.

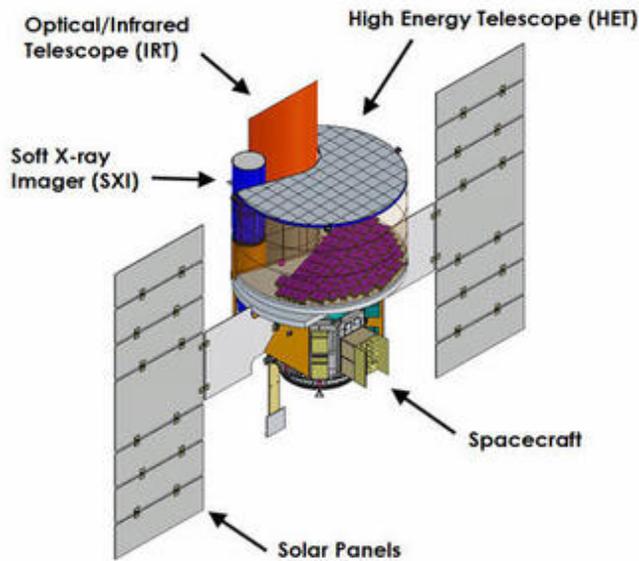

**FIGURE 1.** *EXIST* instruments and layout on spacecraft.

### *EXIST* Instrumentation

The Instrument complement for *EXIST* is shown in Fig. 1 and summarized in Table 1. The HET detector array of imaging CZT (cf. Table 1) is shown as the dark blue array of modules (each with independent anti-coincidence BGO shield underneath and data tagging and control) beneath the hybrid coded aperture mask plane above (grey; with support frame shown by the grid pattern). The IRT (red) is shown with its extended baffle (terminating in the 45° cut) for rejection of scattered sun/moon (45° avoidance angle) and bright earth limb (~20° avoidance). The IRT is surrounded by a thermal shroud (below red baffle) which transfers waste heat from the HET to provide thermal control for the IRT primary/secondary mirror to maintain them at a constant -30C temperature for optimum thermal background (at cosmic zodiacal light levels at ≤2.2μm) while not compromising the mechanical stability of the optical bench and focus control of this commercially developed telescope optical telescope assembly (OTA). The IRT focal plane detectors are derived from the NIRCAM (HIIRG) detectors developed for JWST and are cooled by a cryocooler, The optical side is the HiViSI system incorporating the same HIIRG readout system but with a CMOS detector bumped to the readout instead of the HgCdTe array on the IR side. The SXI, with electroformed thin (0.3mm) mirror shells and significantly lower mass than on XRT despite its greatly enhanced effective area, is provided with a Peltier cooler system to maintain the CCD focal plane at fixed optimal operating temperature.

**TABLE (1).** *EXIST* Instrument and mission parameters (ASMC Design Study)

| Instrument | Telescope & Detector | Energy band | Field of View | Imaging Resol. & Pos. | Spectral Resol. (E/ΔE) | Power (W) | Mass (kg)[1] | Telem. (Mbs)[1] | Heritage |
|---|---|---|---|---|---|---|---|---|---|
| **HET (total Instrument)** | 4.5m² CodedAp | | | | | 726 | 2171 | 4.9 | Swift, INTEGRAL |
| Imager (incl. mask, struct.) | CZT | 5-600 keV | 90° x 70° | 2' & <20" | 5-200 | 341 | 1533 | | Swift/BAT |
| RearShield | BGO | 0.2-2MeV | 2π | N/A | ~10 | 55 | 445 | | INTEGRAL/ IBIS |
| Data proc. | | | | | | 125 | 35 | | |
| Pwr&Therm. | | | | | | 205 | 158 | | |
| **IRT (total Instrument)** | 1.1m Teles. | | | | | 249 | 704 | 1.2 | JWST |
| Imager | HyVISI H2RG | 0.3-0.52μm 0.52-0.9μm 0.9-1.38μm 1.38-2.2μm | 4'x4' | 0.15" | 3 | | | | |
| LoResSpec | Obj.Prism | same | 0.5' | 0.15" | 30 | | | | |
| HiResSpec | LongSlit | same | 10" | 0.15" | 3000 | | | | |
| **SXI (total Instrument)** | 0.6m Wolter-I | | | | | 165 | 313 | 0.08 | Swift/XRT |
| Imager/Spec | CCD | 0.1-10keV | 20' | 15" | 2-60 | | | | |
| **Spacecraft** | | | | | | 1020 | 2147 | 0.02 | Swift, Fermi |
| **TOTALS** | | | | | | 2160 | 5335 | 6.2 | |

NOTES: 1. Includes contingency at 30%

In Figure 2 we show schematic views of the coded aperture telescope for HET and its array of CZT detectors and integrated BGO shields, graded-passive side shields (Hong et al 2009; H09). The coded aperture Tungsten mask is a

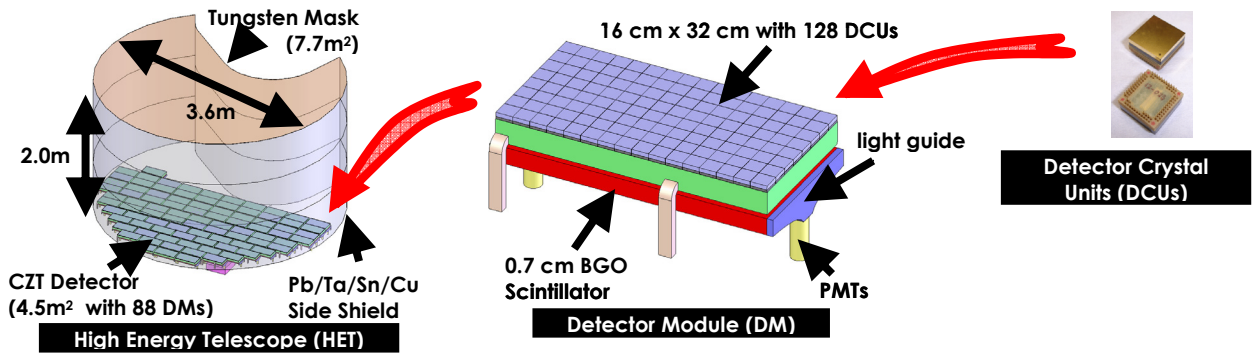

**FIGURE 2.** Schematic layout of the HET showing integration of the 4.5m² focal plane from 88 Detector Modules each consisting of an array of 16 x 8 CZT crystals, each 2 x 2 x 0.5cm and each with 32 x 32 pixels (0.6mm pitch).

Hybrid of coarse (14mm) and fine (1.15mm) holes, both random and each ~50% open, where the fine pattern is etched into a single 0.3mm thick Tungsten sheet that "closes" each open hole of the coarse mask pattern. This 2-scale coded mask (Skinner and Grindlay 1993) provides simultaneous fine resolution (2.4' FWHM) imaging up to ~100keV and coarse resolution (28' FWHM) imaging up to ~600keV and enables rapid (2-step) processing (Skinner et al 2009). Centroiding source positions at <100keV (where sources are most bright) then gives ≤20" source positions (90% confidence radii) for survey threshold (≥5-6σ) sources. GRBs (and all survey sources) are thus located ~100X more precisely than the ≥3' positions now provided by Swift/BAT by image triggers running continuously for both scanning and pointed data.

A schematic of focal plane of the IRT detectors is shown in Figure 3 and the optical path is indicated in Figure 4 (Kutyrev et al. 2009). The IRT incorporates the HIIRG readout which allows non-destructive image readout on short timescales so that imaging S/N can be measured on the fly for objects within the HET error box (shown in Figure 3 as 15" for a GRB near threshold; brighter GRBs are even better located). The sensitive SXI will give an early afterglow position to ≤2" for most GRBs to enable prompt identification of the optical/IR source and constraints on NH and metals in the host galaxy.

### *EXIST* Mission Profile: ~2500 GRB z's in 5y?

How would *EXIST* operate? To achieve maximum (counting-statistics limited) sensitivity, the wide-field HET (and co-aligned IRT and SXI) would be near-zenith pointed on the supporting spacecraft in low earth orbit (LEO; 600km x 15-20° inclination). By pointing ~30° north and then south on alternate orbits, the HET scans nearly the whole sky (except for a 45° avoidance zone around the Sun) every two orbits. On board software is continuously forming images (on a range of timescales, ~1 − 1000sec) from the continuous scan photon event data and onboard ~1" aspect, using rapid coded aperture scanning techniques now developed and demonstrated for analysis of *Swift/BAT* slew survey (BATSS) data (Copete et al 2009). The scanning sky survey is continuous, then, for the HET and SXI (despite its narrow FoV, the full-sky survey at 0.1-10 keV is covered every ~6 months with sensitivity and energy band exceeding ROSAT). When a GRB is detected (on average ~2X/day; see below), the position is computed onboard within ~10sec and a rapid (~100-200sec) slew is initiated to the GRB position provided sun and earth limb constraints are met (with constraints very similar to those for *Swift*). After stable (arcsec) pointing on the GRB position (<20" uncertainty from HET) is achieved, the now on-axis HET is a factor of ~2 more sensitive than in scanning mode over its much larger FoV (where sources are typically only partially coded) so that the high energy coverage of the GRB can continue for (much) longer than with *Swift*/BAT.

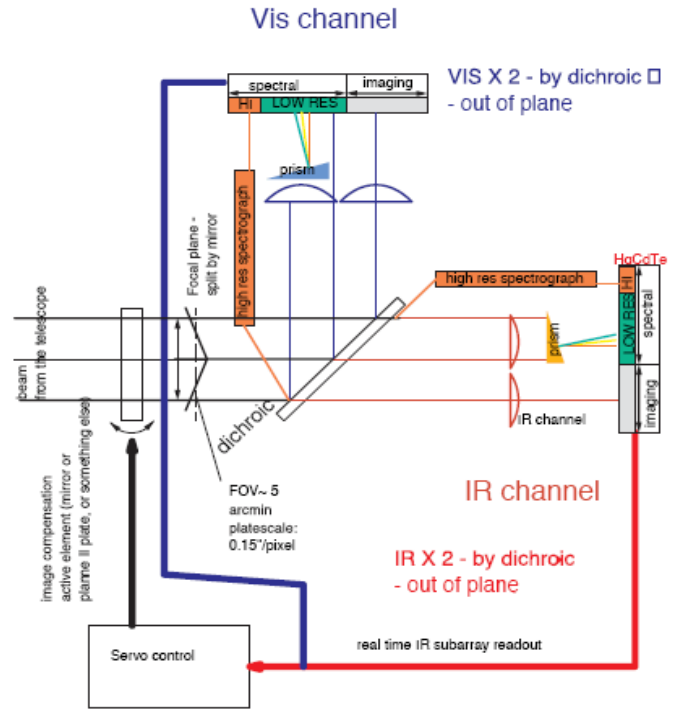

**FIGURE 4.** Optical path in IRT instrument, showing dichroics for simultaneous VIS+IR imaging or spectroscopy and servo image compensation (tip/tilt mirror) for fine guiding.

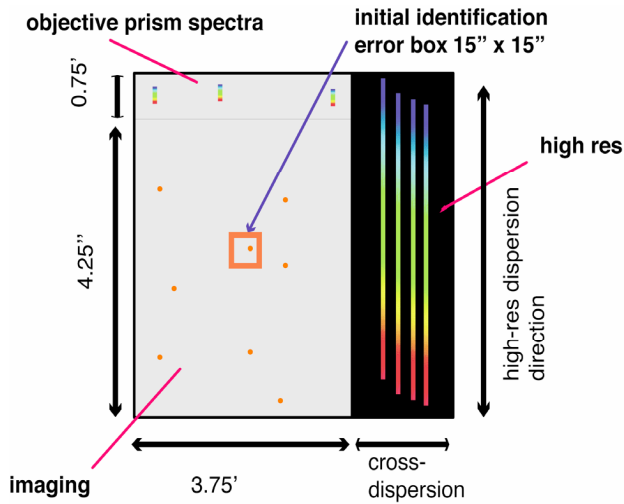

**FIGURE 3.** Imaging vs. spectroscopy allocation in IRT focal plane.

Initial SXI and IRT images and colors are obtained during the first 100sec of pointing that further refine (from the SXI) the GRB position to <2" to allow prompt GRB identification in the IRT data. A 30" window in each of the 4-band IRT images obtained simultaneously via the dichroics (Table 1) is processed on board to find either a "new" object not present on a stored (on board) equivalent of the DSS catalog (compressed tables only) or by measured variability of any object within the SXI <2" error box. High redshift GRBs will in many/most cases be immediately identified by dropouts in the blue band images. With GRB identification then obtained typically by ~300sec after trigger, if the object is brighter than AB <20, it is centered on the hi-res (R = 3000) spectrograph long slit by either a small S/C pointing adjustment (if offset by >10" from the initial HET position) or, if closer, by the tip/tilt guiding mirror within the IRT instrument (see Fig. 4). For most GRBs, the hi-res spectra and redshifts will be possible for at least the first ~1000sec, allowing high resolution measures of absorption line spectra for DLA and EOR studies and metalicities in the host as well as IGM absorption features. If the object is fainter than AB ~20 in the "detection" bands, it is instead moved onto the objective prism portion of the detector FoV for a lo-res (R = 30) spectrum to measure the GRB redshift and absorption (or emission) line spectrum. For either spectrograph selection, spectra are obtained simultaneously in all 4 bands for complete coverage at fixed resolution over the full 0.3 – 2.2μm band. Throughout either initial imaging or followup spectra, fine pointing is achieved by the servo tip/tilt mirror to maintain <0.1" pointing stability on a S/C that need only point to ~3", allowing 0.15" pixel images in all 4 bands.

Most GRBs would be followed for spectroscopy for two consecutive orbits (with scanning resumed or a pointing on the most recent prior GRB on the "other side" of each orbit). The near-zenith-scan survey then resumes until the next GRB or transient (of unusual nature) triggers the next slew-pointing, for a total of ~4 orbits of pointing each day for the first 2y of the mission. During the final 3y of the mission, *EXIST* is primarily a pointed mission – following up on the thousands of new sources (AGN, unusual transients) discovered in the scanning sky survey, and still continuing to trigger on GRBs at the same anticipated rate (~2 per day). Given Poisson arrival times, there will often be 4-5 GRBs per day(!) so that on-board decision making will be made to prioritize which GRB to resume pointing on for additional spectroscopy vs. resuming full-sky scanning. First-order prioritization will, of course, go the currently highest redshift GRB with magnitude still bright enough (at AB <21-22) for hi-res spectroscopy in typical ~2000sec integrations possible on a given orbit, or for stacked spectra over several orbits.

The sky survey (scanning) sensitivity for the HET, averaged over its FoV (H09), is such that a GRB with duration T90 ~3-10s and a typical power law spectrum with photon index 1 – 2 can be detected (≥10σ) and located (≤10") at fluence $S = 5 \times 10^{-9}$ erg/(cm$^2$ –sec) in multiple bands across the 5-100keV band. This is a factor of ~10 below the faintest long burst, GRB080520 with T90 = 2.8s, listed on the BAT GRB table, http://swift.gsfc.nasa.gov/docs/swift/archive/grb_table/ as of March 1, 2009. Given that the *EXIST/HET* field of view for GRBs (10% coding: 90° x 70°) vs. that for *Swift/BAT* (105° x 80°) is smaller by a factor of 1.3, and the logN-logS slope of -0.8 for the faint end of the *Swift* GRB sample (Dai et al 2008), the predicted GRB rate for *EXIST* is expected to be a factor of ~6 larger than the rate for *Swift*, or ~600 GRBs/year. Thus over a 5y mission lifetime, some 3000 GRBs and very likely >2500 redshifts are possible vs. a projected total of ~200 redshifts over 5y from *Swift*. Given predictions (BL07, Salvaterra et al 2008) that >10% of *EXIST* GRBs will be at z >6, and allowing for the higher overall sensitivity, extended low energy GRB trigger capability and – primarily – the on board IRT, this 10% figure is probably conservative. The number of sight-lines "drilled" by *EXIST* GRBs in the z >6 Universe should be at least ~250, allowing statistically significant probes of the evolution of structure and the EOR.

## WHAT GRBS (AND MORE) COULD REVEAL TO *EXIST*…

As a powerful multi-wavelength observatory, *EXIST* is designed to detect and study high-z GRBs to use these flashbulbs on the early Universe to reveal its structure and evolution. In Figure 5 we present simulated IRT spectra for what would be obtained from what might be a "typical" GRB at redshift *greater* than the currently highest known, the z = 6.7 GRB080913, which was itself some 4mag fainter (at ~1000sec after GRB trigger) than the second largest redshift GRB050904 at z = 6.4. The spectra were simulated by J. Bloom (using IRT sensitivities and resolutions) and are from McQuinn et al (2009). The spectra shown are (on left) what the low resolution (R = 30 spectrograph on the IRT would provide for a 1000sec integration, after acquisition, on a GRB one mag. *fainter* than this GRB and redshifted to the values shown. These lo-res spectra are shown across all 4 bands. On the right are spectra (plotted in rest frame) scaled from the same GRB080913 but now 2 mag brighter, to be more typical, and now observed with the hi-res spectrograph with R = 3000. For both spectra, the GRB spectrum is assumed to begin at T = 300s after GRB trigger and then decay with typical decay $F(\tau) \sim \tau^{-1}$. The host galaxy is assumed to have local column density, log(NH) =20, and the absorption line spectrum (shown) is governed by the assumed metallicity vs. z, namely: [Fe/H] = -2 (z <6), [Fe/H] = -3 (6 < z < 7), and [Fe/H] = -5 (z >7). As can be seen, GRB redshifts can be readily measured in the first 1000sec down to AB(mag) ~25, and narrow line spectral identifications are possible down to AB ~22. The SXI spectrum (~140 eV resolution) will itself constrain metals in the host and thus metallicity in the early Universe, as was done for the GRB050904 at z = 6.3 (Campana et al 2007). The key to GRBs as cosmological probes, then, is this prompt measurement across a broad band into the NIR when GRBs are sometimes even brightening (flares in some early afterglows) and thus most readily detected. The passively cooled (-30C) telescope reduces background at 2μm a factor ~$10^3$ below ground-based IR telescopes to yield the speed advantage (~10X faster than Keck at H band) and sensitivities shown.

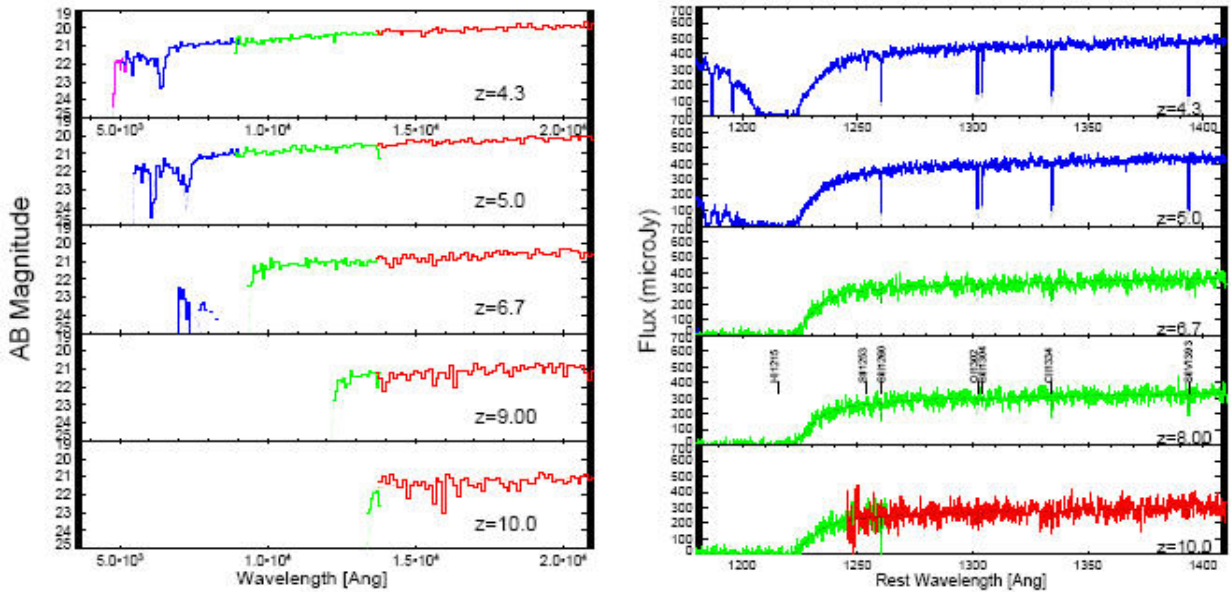

**FIGURE 5:** *(left)* Simulated IRT spectra with the R = 30 spectrograph for a GRB 1mag fainter (at AB ~21) than GRB080913 (z =6.7), among the faintest measured for *Swift*; *(right)* same GRB but a more typical 2mag brighter and observed with the proposed R =3000 spectrograph. Red, green, blue and purple (*see electronic version*) show the 4 contiguous spectral bands.

The high sensitivity of the HET, SXI and IRT also enable qualitatively new GRB physics to be explored. The HET can measure polarization in GRBs at levels ≤20% for bright GRBs with fluence ≥1 x $10^{-6}$ ergs/cm$^2$ (or persistent sources such as Blazars with flux ≥10mCrab over the 5y mission) in the ~80-300 keV band by its fine-pixel imaging sensitivity as a Compton polarimeter (Krawcynski et al 2009). GRB spectra on short timescales (~1msec) and with ~3X the resolution and bandwidth of BAT will allow new constraints on the nature of GRB jets and whether they have fine structure as has been proposed, while the SXI will enable the physics of puzzling flares in early afterglows (particularly for short GRBs) to explored in much greater detail and simultaneous with the initial broad-band IRT imaging.

Finally, as described in recent Astro2010 White Papers, *EXIST* opens Discovery space in two major other areas: studies of supermassive black holes (accreting, obscured, and dormant) in the deepest full-sky hard X-ray imaging survey possible (Coppi et al 2009); and studies of stellar and intermediate mass black hole Transients (Grindlay et al 2009) and, more generally, the time-variable Universe triggered by HET (>5 keV) and ranging from stellar flares throughout the Galaxy, to supernovae in the Local Group, and both magnetar giant flares and tidal distruption flares (by quiescent supermassive black holes) out to ~300Mpc (Soderberg et al 2009).

## ACKNOWLEDGMENTS

I thank the GRB Science and Instrument Technical Working Groups and their chairs for considerable progress in developing the science and technical study for *EXIST*. The GRBSWG (Josh Bloom), and in particular Matt McQuinn,; the HETTWG (Jaesub Hong) and Scott Barthelmy; the HET Imaging TWG (Gerry Skinner); the IRTTWG (Harvey Moseley) and Alexander Kutyrev; and the S/C and mission TWG (Jerry Fishman) and Dom Conte – have all contributed enormously. This work is supported in part by the ASMC Study grant, NNX08AK84G.